\documentclass[onecolumn, nofootinbib, pra]{revtex4} 
\usepackage{amssymb, amsmath}
\usepackage{graphicx}
\usepackage{verbatim}
\usepackage{xcolor}
\begin{document}
\title{Reply to superdeterminists on the hidden-variable formulation of Invariant-set theory}
\author{Indrajit Sen}
\email{isen@chapman.edu}
\affiliation{Institute for Quantum Studies, Chapman University\\
One University Drive, Orange, CA, 92866, USA}
\date{\today}

\begin{abstract}
In a recent article (arXiv:2108.08144), Hance, Hossenfelder and Palmer have advanced arguments claiming to show that the analysis of Invariant-set theory in a hidden-variable setting (arXiv:2107.04761) is wrong. We explain here why these arguments are incorrect, and involve misunderstandings of the hidden-variable model in question and Bell's notion of local causality.
\end{abstract}
\maketitle

\section{Introduction}
Superdeterminism posits a deterministic universe with such initial conditions that the measurement settings and the hidden variables (that determine the measurement outcomes) in Bell experiments are correlated \cite{dialect}. Superdeterministic hidden-variable models thereby circumvent Bell's theorem \cite{bell} by violating the measurement-independence assumption \cite{howmuch, hall10}, and can give a local account of the Bell correlations. However, superdeterminism has a conspiracy problem. This was pointed out by Bell, who gave a short qualitative argument to that effect \cite{bertlmann}. Recent work has shown that the notion of superdeterministic conspiracy can be made quantitative in two separate ways \cite{1st, 2nd}. \\

Palmer has proposed \cite{palmer09, palmer1, palmer2, palmerend, palmerpost} a superdeterministic account of quantum mechanics and claimed it to be local, $\psi$-epistemic \cite{harrikens, leifer} and free of the conspiracy problem. However, it is difficult to evaluate any of the claims without a concrete hidden-variable model. Recently, we attempted to fill this gap by building a hidden-variable model based on Palmer's proposal and then analysing it. We found that the resultant model is nonlocal with an ontic quantum state, and that several arguments that appear to work in the proposal fail when considered in the model \cite{senysis}. We also showed that the model is conspiratorial. Palmer and co-workers (referred to as ``the authors" hereafter for convenience) have responded by providing arguments to claim that the model does not accurately represent the proposal \cite{totalpagol}. \\

Evaluating this claim opens up a question of principle: how do we know whether a model, based on a proposal, is an inaccurate representation of the proposal? The difference in clarity between a mathematical model and a proposal for a model can make any comparison a difficult task. We suggest that this question be answered by considering the objections raised against the accuracy of the model. Are these objections correct? If not, then one cannot conclude that the model is an inaccurate representation of the proposal. Evaluating the objections is straightforward as a hidden-variable model, unlike a proposal for such a model, can be transparently analysed. In this article, we show that the objections raised in ref. \cite{totalpagol} against our model are incorrect. We analyse the different arguments by the authors in section \ref{II}, and briefly conclude in section \ref{III}. It is assumed that the reader has some familiarity with ref. \cite{senysis}.\\

\section{Response to objections} \label{II}

\begin{figure}
\includegraphics[scale=0.4]{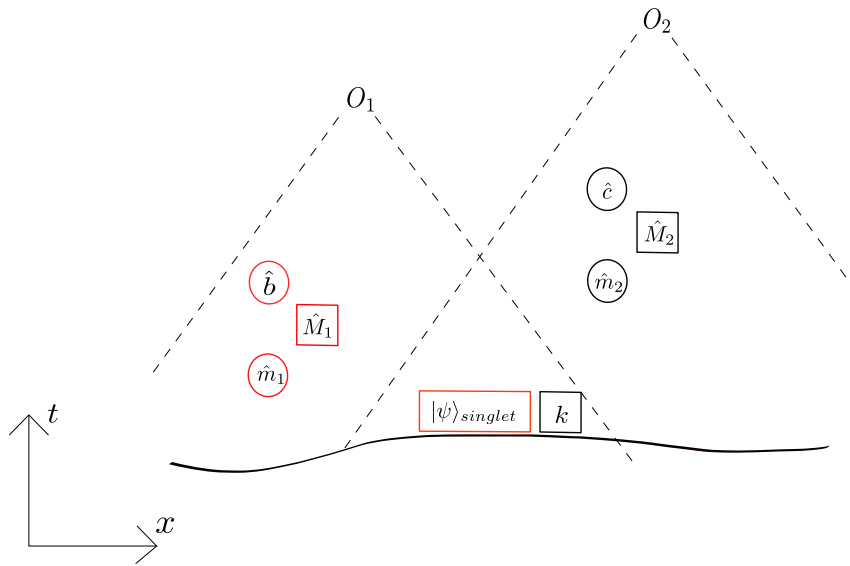}
\caption{Schematic illustration of the nonlocality in the model. The outcome at the second wing of the Bell experiment $O_2$ depends on the hidden variable $\mu \equiv \{\hat{M}_1, \hat{M}_2, |\psi\rangle_{singlet}, k\}$ and the experimenters' choices $\hat{m}_1$, $\hat{b}$ at wing 1 and $\hat{m}_2$, $\hat{c}$ at wing 2. The hidden variables (experimenters' choices) are boxed (circled) in the figure. The model violates Bell's notion of local causality as the outcome $O_2$ is determined by the variables $\hat{m}_1$, $\hat{M}_1$ and $\hat{b}$ that are space-like separated w.r.t. $O_2$, and by $|\psi\rangle_{singlet}$ which is associated with spacelike hypersurfaces belonging to a foliation of spacetime. These variables are highlighted in red. The variable $\hat{m}_1$ ($\hat{b}$) labels the initial (final) experimentally-selected orientation of the Stern Gerlach at the first wing for a particular run. The variable $\hat{M}_1$ labels the exact orientation of the Stern Gerlach at the first wing before the experimenter selects $\hat{b}$. }
\end{figure}
\subsection{On discreteness}
The authors claim that our model of the proposal makes an error in discretizing the Bloch sphere. First, they argue that we have misunderstood how ``the discrete grid of valid points (formed by the intersection of the $N$ lines of latitude and $N$ lines of longitude allowed) forces eigenstates measured to the closest point to the eigenstate we'd expect for the given operator measured (e.g. eigenstates of $\hat{P}$ instead of those of $\hat{p}$), but, depending on the idealised operator measured, the change between the ideal and experimentally-allowed eigenstate has no reason to have to be the same for different operators.'' \cite{totalpagol}. However, there are no eigenstates corresponding to the vectors $\hat{P}$ and $\hat{p}$ in our model. Actually $\hat{P}$ ($\hat{p}$) is the exact (experimentally-selected) orientation of the Stern Gerlach \textit{before} the experimenter has selected the final experimentally-selected orientation $\hat{a}$ for that run. Second, they argue that, in our model ``there is no reason $\delta\hat{A}$ need equal $\delta\hat{P}$'' \cite{totalpagol}. However, we have not used the relation $\delta\hat{A} = \delta\hat{P}$ in the first place. The relation actually used is $|\delta\hat{A}(\hat{P}, \hat{p}, \hat{a})| = |\delta\hat{P}|$. Furthermore, the relation $|\delta\hat{A}(\hat{P}, \hat{p}, \hat{a})| = |\delta\hat{P}|$ is not necessary for our model to work. It is a consequence of the specific transformation considered -- for illustration -- between the initial and final exact orientation of the Stern Gerlach given the experimenter's choices $\hat{p}$ and $\hat{a}$. Our results hold for \textit{any} such transformation given that it satisfies $|\delta\hat{P}| < \Delta \Rightarrow |\delta\hat{A}(\hat{P}, \hat{p}, \hat{a})| < \Delta$ (and conditions of linearity and invertibility) \cite{senysis}. 

\begin{figure}
\includegraphics[scale=0.4]{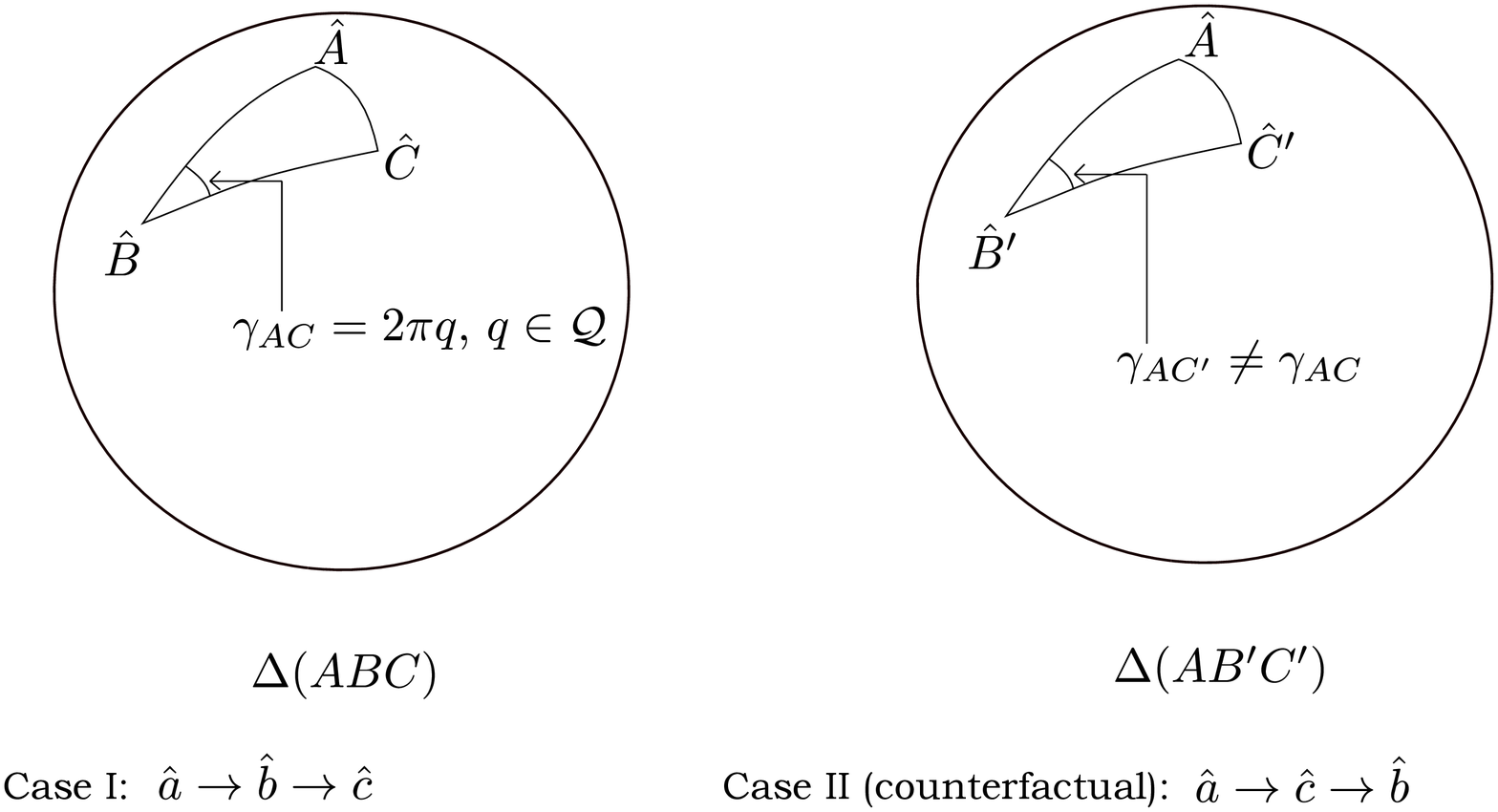}
\caption{Schematic illustration of the application of Niven's theorem (a number-theoretic result) in the model. The left-hand side of the figure depicts a scenario where the experimenters have chosen the experimentally-selected settings to be $\hat{a} \to \hat{b} \to \hat{c}$ for a particular run of a sequential Stern-Gerlach measurement. The corresponding exact measurement settings $\hat{A}$, $\hat{B}$ and $\hat{C}$ for that run form the spherical triangle $\Delta(ABC)$ on the unit sphere. The proposal requires that $\gamma_{AC}/2\pi \in \mathcal{Q}$, where $\mathcal{Q}$ is the set of rational numbers. This implies, via Niven's theorem, that $\hat{B}\cdot \hat{C} \not \in \mathcal{Q}$ (except for certain special cases). The right-hand side of the figure depicts a counter-factual scenario for that run, where the order of experimentally-selected settings is $\hat{a} \to \hat{c} \to \hat{b}$, and the times at which the measurements occur are the same as before. For this counter-factual scenario, the exact measurement settings turn out to be $\hat{A}$, $\hat{B}'$ ($ \neq \hat{B}$ in general) and $\hat{C}'$ ($\neq \hat{C}$ in general) in the model, and the relevant spherical triangle is $\Delta(AB'C')$. For the $\Delta(AB'C')$,  $\gamma_{AC'}/2\pi \not \in \mathcal{Q}$ in general as $\gamma_{AC'} \neq \gamma_{AC}$. Whether such a counter-factual scenario could have occurred for that run depends on whether $\gamma_{AC'}/2\pi$ is rational.}
\end{figure}

\subsection{On nonlocality}
The authors give two arguments why they consider our analysis of nonlocality to be wrong. First, they argue that ``in a superdeterministic hidden variables theory, the full specification of the prepared state already contains information about the measurement settings. The theory then trivially fulfils local causality because it is unnecessary to add the measurement settings a second time to calculate the measurement outcome'' \cite{totalpagol}. There are multiple problems with this argument. First, as the measurement outcomes depend on the hidden variables and the measurement settings, it is unnecessary to add the measurement settings (as the authors claim) only if the hidden variables that determine the outcomes already contain \textit{complete} information about the measurement settings. However, this does not follow from the definition of a superdeterministic model given by two of the authors (Hossenfelder and Palmer) in an earlier work. In ref. \cite{sabinethink}, they define a superdeterministic model ``as a Psi-epistemic, deterministic theory that violates Statistical Independence but is local in the sense of respecting Continuity of Action''. The conjunction of these properties does not imply that the ``full specification of the prepared state'' must contain complete information about the measurement settings. Furthermore, a model where the hidden variables that determine the measurement outcomes contain complete information about the measurement settings is not trivially locally causal. Bell defined local causality in the following manner:
\begin{quote}
``A theory will be said to be locally causal if the probabilities attached to values of local beables in a space-time region 1 are unaltered by specification of values of local beables in a space-like separated region 2, when what happens in the backward light cone of 1 is already sufficiently specified, for example by a full specification of local beables in a space-time region 3 [located in the backward light cone of 1 or in the overlap of backward light cones of 1 and 2].'' \cite{cuisine}
\end{quote}
Therefore, for the Bell scenario, if the hidden variables that contain the measurement settings at the first wing are not localised in the backward light cone of the measurement outcome event at the second wing, then the model is not locally causal. This can be clearly illustrated by considering pilot-wave theory \cite{bohm1, bohm2, hollandbook, solventini}, which is a deterministic theory of the entire universe. In this theory, the measurement settings chosen by an experimenter are, in principle, determined by the hidden variables that describe the experimenter and the physical factors that influence their choice. However, it would be incorrect to conclude from this that pilot-wave theory is locally causal: the hidden variables that determine the measurement settings (the positions of the particles constituting, and the quantum states of, the experimenter and the physical factors) at one wing of a Bell experiment are not localised in the backward light cone of the measurement-outcome event at the other wing. \\

We now explain why our model violates Bell's notion of local causality. Consider, in our model for the Bell scenario, the measurement outcome at the second wing $O_2$ as a function of the hidden variable $\mu= \big ( |\psi\rangle_{singlet}, \hat{M}_1, \hat{M}_2, k \big )$ and the experimenters' choices $\hat{m}_1$, $\hat{b}$ ($\hat{m}_2$, $\hat{c}$) at the first (second) wing. Here $\hat{M}_1$ ($\hat{M}_2$) labels the exact orientation of the Stern-Gerlach apparatus at the first (second) wing before the experimenter at the first (second) wing has selected $\hat{b}$ ($\hat{c}$) for that run. The outcome can be represented as $O_2(\mu, \hat{b}, \hat{m}_1, \hat{c}, \hat{m}_2)$ \cite{senysis}. As $O_2$ depends on $\hat{m}_1$, $\hat{b}$, $\hat{M}_1$ and $|\psi\rangle_{singlet}$, the model violates local causality in the sense of Bell. This is illustrated in Fig. 1. \\

Second, the authors repeat the claim made in the proposal that any counter-factual change in the measurement setting at one wing while keeping the hidden variables and the measurement setting at the other wing constant leads to a violation of a rationality constraint imposed by the proposal. However, we have shown that this argument fails to work in our model of the proposal \cite{senysis} (we discuss their claims on counterfactuals in section \ref{ad}). 

\subsection{On $\psi$-onticity}
The authors claim that our analysis that the model is $\psi$-ontic is wrong. However, they argue why the bit string, instead of the quantum state, is not a hidden variable\footnote{``This statement is incorrect because the bit-string is not the ontic state of the model, it is an ensemble of ontic states.'' \cite{totalpagol}}. But the bit string is not identical with the quantum state: the former contains information about both the exact quantum state prepared \textit{and} the exact final measurement setting for a particular run. \\

The model is $\psi$-ontic as the measurement outcome for any run of the experiment depends on the exact quantum state prepared for that run. In a $\psi$-epistemic model, on the other hand, the quantum state is relevant only as a statistical description at the ensemble level (not at the level of the individual run). The bit string is also real in the model: it is determined by the exact quantum state prepared and the exact final measurement setting for a particular run, both of which are hidden variables that determine the measurement outcome for that run. \\

We note here that the the usual distinction between hidden variables and measurement settings cannot be maintained in a superdeterministic model where the exact measurement settings are themselves hidden variables. This does not, however, introduce any ambiguity about the ontological status of a variable: if the variable plays a role in determining the measurement outcome for an individual run of the experiment, then that variable is ontologically real in the model.

\subsection{On counterfactuals} \label{ad}
The authors claim that our model fails to use the number-theoretic results in the proposal that are important in evaluating counter-factual scenarios\footnote{``Sen never invokes the number-theoretic incompatibility between angles and their cosines in his paper.''\cite{totalpagol}}. However, the claim is incorrect. In our model, whether a counter-factual scenario for a particular run is possible or not depends on the number-theoretic result, Niven's theorem \cite{niven}, applied to the spherical triangle appropriate to the counter-factual scenario for that run.\\

Where our analysis actually differs from the proposal is in the specific spherical triangle to which Niven's theorem in applied (see Fig. 2). In general, the spherical triangle for the counter-factual scenario is different from the spherical triangle for the actual scenario, due to the following reason. Suppose that, for a particular run of a sequential Stern-Gerlach measurement, the experimentally-selected orientation of the second (third) Stern-Gerlach apparatus is $\hat{b}$ ($\hat{c}$). Let the second (third) measurement be performed at time $t_2$ ($t_3$), where $t_3 > t_2$. As the exact apparatus orientations are time dependent, the exact orientation of the second (third) Stern-Gerlach apparatus at the time of measurement may be labelled by $\hat{B}(t_2)$ $\big (\hat{C}(t_3)\big )$. The order of experimentally-selected (exact) settings will be $\hat{b} \to \hat{c}$ $\big ( \hat{B}(t_2) \to \hat{C}(t_3) \big )$ for this run. Consider a counter-factual scenario for the run where the order of Stern-Gerlach apparatuses is reversed, keeping the experimentally-selected orientations and the measurement times fixed. That is, the second (third) Stern-Gerlach apparatus is used to measure the quantum particle at $t_3$ ($t_2$). The order of experimentally-selected orientations will then be $\hat{c} \to \hat{b}$, and the order of exact orientations will be $\hat{C}(t_2) \to \hat{B}(t_3)$, for this counter-factual scenario. Therefore, a counter-factual reversal in the order of experimentally-selected settings does not, in general, imply that the same exact settings are reversed. \\

On the other hand, the proposal makes an implicit, and physically implausible, assumption that $\hat{B}(t_2) = \hat{B}(t_3)$ and $\hat{C}(t_2) = \hat{C}(t_3)$. It then claims to rule out, via Niven's theorem, a counter-factual change in the order of experimentally-selected settings\footnote{``Therefore, if $U$ is a universe in which $\hat{a} \to \hat{b} \to \hat{c}$ is performed on a particular particle, and therefore lies on $I_U$, the counterfactual universe $U'$ where $\hat{a} \to \hat{c} \to \hat{b}$ is performed on the same particle cannot lie on $I_U$.''\cite{palmerend}}. However, this is incorrect: it only rules out the counter-factual scenario where both the experimentally-selected \textit{and} the same exact measurement settings are reversed in order. Lastly, we note that the proposal makes the same assumption while considering counter-factual scenarios in the Bell setup.\\ 

\section{Conclusion}\label{III}
We have shown that the arguments made by the authors regarding our model are incorrect. Therefore, their claim that the model does not accurately represent Invariant-set theory cannot be supported on the basis of their arguments.\\

\acknowledgements
I am thankful to Matthew Leifer for useful discussions. I am also thankful to Tim Palmer, Jonte Hance and Sabine Hossenfelder for discussing their work. I acknowledge support by the GCI fellowship at Chapman university.
\bibliographystyle{bhak}
\bibliography{bib}

\begin{thebibliography}{10}

\bibitem{dialect}
J.~S. Bell, A.~Shimony, M.~A. Horne, and J.~F. Clauser.
\newblock {An exchange on local beables}.
\newblock {\em Dialectica}, pages 85--110, 1985.

\bibitem{bell}
J.~S. Bell.
\newblock {\em {Speakable and unspeakable in quantum mechanics: Collected
  papers on quantum philosophy}}.
\newblock {Cambridge Univ. Press}, {2004}.

\bibitem{howmuch}
J.~Barrett and N.~Gisin.
\newblock {How much measurement independence is needed to demonstrate
  nonlocality?}
\newblock {\em Phys. Rev. Lett.}, 106(10):100406, 2011.

\bibitem{hall10}
M.~J. Hall.
\newblock {Local deterministic model of singlet state correlations based on
  relaxing measurement independence}.
\newblock {\em {Phys. Rev. Lett.}}, {105}({25}), {2010}.

\bibitem{bertlmann}
J.~S. Bell.
\newblock {Bertlmann's socks and the nature of reality}.
\newblock In {\em {Speakable and unspeakable in quantum mechanics: Collected
  papers on quantum philosophy}}. {Cambridge Univ. Press}, {2004}.

\bibitem{1st}
I.~Sen and A.~Valentini.
\newblock {Superdeterministic hidden-variables models I: nonequilibrium and
  signalling}.
\newblock {\em Proc. R. Soc. A}, 476(2243):20200212, 2020.

\bibitem{2nd}
I.~Sen and A.~Valentini.
\newblock {Superdeterministic hidden-variables models II: conspiracy}.
\newblock {\em Proc. R. Soc. A}, 476(2243):20200214, 2020.

\bibitem{palmer09}
T.~Palmer.
\newblock {The Invariant Set Postulate: a new geometric framework for the
  foundations of quantum theory and the role played by gravity}.
\newblock {\em Proc. R. Soc. A}, 465(2110):3165--3185, 2009.

\bibitem{palmer1}
T.~Palmer.
\newblock {Lorenz, Godel and Penrose: new perspectives on determinism and
  causality in fundamental physics}.
\newblock {\em Contemp. Phys}, 55, 2014.

\bibitem{palmer2}
T.~Palmer.
\newblock {Bell's conspiracy, Schrodinger's black cat and global invariant
  sets}.
\newblock {\em Phil. Trans. R. Soc. A}, 373, 2015.

\bibitem{palmerend}
T.~Palmer.
\newblock {Discretization of the Bloch sphere, fractal invariant sets and
  Bell’s theorem}.
\newblock {\em Proc. R. Soc. A}, 476(2236):20190350, 2020.

\bibitem{palmerpost}
J.~R. Hance, T.~N. Palmer, and J.~Rarity.
\newblock {Experimental Tests of Invariant Set Theory}.
\newblock {\em arXiv:2102.07795}, 2021.

\bibitem{harrikens}
N.~Harrigan and R.~W. Spekkens.
\newblock {Einstein, incompleteness, and the epistemic view of quantum states}.
\newblock {\em {Found. Phys.}}, {40}({2}), {2010}.

\bibitem{leifer}
M.~S. Leifer.
\newblock {Is the quantum state real? An extended review of $\psi $-ontology
  theorems}.
\newblock {\em {arXiv:1409.1570}}, {2014}.

\bibitem{senysis}
I.~Sen.
\newblock {Analysis of the superdeterministic Invariant-set theory in a
  hidden-variable setting}.
\newblock {\em arXiv:2107.04761}, 2021.

\bibitem{totalpagol}
J.~Hance, S.~Hossenfelder, and T.~Palmer.
\newblock {Comment on ``Analysis of the superdeterministic invariant-set theory
  in a hidden-variable setting"}.
\newblock {\em arXiv:2108.08144}, 2021.

\bibitem{sabinethink}
S.~Hossenfelder and T.~Palmer.
\newblock {Rethinking superdeterminism}.
\newblock {\em Frontiers in Physics}, 8:139, 2020.

\bibitem{cuisine}
J.~S. Bell.
\newblock {La nouvelle cuisine}.
\newblock In {\em {Speakable and unspeakable in quantum mechanics: Collected
  papers on quantum philosophy}}. {Cambridge Univ. Press}, {2004}.

\bibitem{bohm1}
D.~Bohm.
\newblock {A suggested interpretation of the quantum theory in terms of
  ``hidden" variables. I}.
\newblock {\em {Phys. Rev.}}, 85(2), 1952.

\bibitem{bohm2}
D.~Bohm.
\newblock {A suggested interpretation of the quantum theory in terms of
  ``hidden" variables. II}.
\newblock {\em {Phys. Rev.}}, {85}({2}), {1952}.

\bibitem{hollandbook}
P.~R. Holland.
\newblock {\em {The quantum theory of motion: an account of the de Broglie-Bohm
  causal interpretation of quantum mechanics}}.
\newblock {Cambridge Univ. Press}, 1995.

\bibitem{solventini}
G.~Bacciagaluppi and A.~Valentini.
\newblock {\em {Quantum theory at the crossroads: reconsidering the 1927 Solvay
  conference}}.
\newblock {Cambridge Univ. Press}, 2009.

\bibitem{niven}
I.~Niven.
\newblock {Carus Math. Monogr., vol. 11}, 1956.

\end{thebibliography}

\end{document}